\documentstyle[12pt]{article}
\begin{document}
\begin{center}{\large{\bf{A note on the vector Schwinger Model with
the photon mass term: Gauge invariant reformulation, operator
solution and path integral formulation}}}\end{center}
\begin{center} Anisur Rahaman \\ Durgapur Govt. College, Durgapur -
712314,
Burdwan, West Bengal, India\\
At present\\ Govt. College of Engineering and Textile Technology,
Serampore-712202\\ Hooghly,West Bengal, India\\
e-mail: anisur.rahman@saha.ernet.in \end{center} \vspace{2cm}

\centerline{Abstract} Vector Schwinger model is reinvestigated
with the mass like term for gauge field. Phase space structure has
been determined in this situation. It has been found that mass of
the gauge boson acquires a generalized expression with the bare
coupling constant and the parameters involved in the masslike
terms for the gauge field which is in contrast with the result of
a recent publication on this issue.
\newpage



QED in $(1+1)$ dimension, e.g., Schwinger model \cite{SCH} is a
very interesting field theoretical model. It has been widely
studied over the years by several authors in connection with the
mass generation, confinement aspect of fermion (quark), charge
shielding etc. \cite{LO, COL, CAS, AG}. The description of this
model in noncommutative space time has also been found to give
interesting result \cite{APR, APR1}. Effect of non commutativity
in space time has showed up as an interacting background with the
massive boson of the usual vector Schwinger model. In the
Schwinger model massless fermion interact with the Abelian gauge
field. Photon acquires mass via a kind of dynamical symmetry
breaking and the quarks disappear from the physical spectra. The
exactly solvable nature of the model leads to express this model
in terms of canonical boson field. It is a remarkable feature of
(1+1) dimensional exactly solvable interacting fermionic field
theory. Gauge invariance was there in the fermionic version of the
model. So initially the model in the bosonised version too studied
maintaining gauge invariance. However gauge non invariant
regularized version of this model also found to be an important
field theoretical model \cite{AR}. Here we find that a one
parameter class of regularization commonly used to study the
chiral Schwinger model has been introduced in the vector Schwinger
model. For a specific choice,
 i.e., for the vanishing value of the parameter the model reduces to
 the usual vector Schwinger model but for
 the other admissible value of this parameter the phase space structure
 as well as  the the physical spectra
gets altered remarkably. This new regularization leads to a change
in the confinement scenario of the quark too. In fact, the quarks
gets liberated as it was happened in the Chiral Schwinger model
\cite{JR, RABIN, GIR, GIR1, GIR2}.

Recently, we find that the Schwinger model is studied adding
masslike term for gauge field with the lagrangian at the classical
level \cite{USHA1, USHA2}. This model is structurally equivalent
to the model studied in \cite{AR}. The masslike term for gauge
gauge field are introduced in the two models with different
perspective. In \cite{AR}, masslike term occurred as a one loop
correction in order to remove the divergence of the fermionic
determinant appeared during bosonization whereas in \cite{USHA1,
USHA2}, the author studied the Schwinger model with the masslike
term for gauge field at the classical level. In \cite{AR}, the
authors had reasonably fair  motivation to introduce the masslike
term since it is known that in QED, a regularization gets involved
when one calculates the effective action by integrating the
fermions out. The ambiguity in the regularization has been
exploited by different authors in different times in (1+1)
dimensional QED and Chiral QED and different interesting scenarios
have been resulted in \cite{JR, RABIN, GIR, GIR1, GIR2, PM, MG,
KH, ABD}. The most remarkable one is the chiral Schwinger model
studied by Jackiw and Rajaraman \cite{JR}. They saved the long
suffering of the chiral generation of the Schwinger model due to
Hagen \cite{HAG} from the non-unitary problem introducing a one
parameter class of regularization. In \cite{CASA}, the authors
studied the Schwinger model introducing masslike term for the
gauge field at the classical level and showed that for a
particular value of the ambiguity parameter the lost gauge
invariance of the so called nonconfining (anomalous) Schwinger
model \cite{AR} gets restored. In \cite{USHA1, USHA2}, however the
authors presented a surprising and untrustworthy result adding the
same masslike term at the classical level. There we find that the
mass generated for the boson is $m=\sqrt{2}e$ and it does not
contain the parameter involved within the masslike term of the
gauge field!
If we look into the the work \cite{AR}, a structurally equivalent
model to \cite {USHA1}, we find that the theoretical spectrum
contains a massive and a massless boson and the mass of the
massive boson acquires a generalized expression with the ambiguity
parameter. However a massless boson is there in \cite{USHA1,
USHA2}. A parameter free mass term may appear if the added term
works as gauge fixing. However the added term was not a legitimate
gauge fixing term in \cite{USHA1, USHA2}. A question therefore,
automatically comes how did the authors get such untrustworthy
result in \cite{USHA1, USHA2}? The model is thus reinvestigated in
this note.

The vector Schwinger model is described by the following
generating functional.
\begin{equation} Z[A] = \int
d\psi d\bar\psi e^{\int d^2x{\cal L}_f}
\end{equation}\b with
\begin{equation}
{\cal L}_f = \bar\psi\gamma^\mu[i\partial_\mu + e\sqrt\pi
A_\mu]\psi \nonumber
\end{equation}
The effective bosonized lagrangian density obtained by integrating
out the  the fermion one by is
\begin{equation}
{\cal L}_B = \frac{1}{2}\partial_\mu\phi \partial^\mu\phi -
e\epsilon_{\mu\nu}\partial^\nu\phi A^\mu. \label{BLD}
\end{equation}
If electromagnetic background is introduced with masslike term for
the gauge field the model reads
\begin{equation}
{\cal L}_B = \frac{1}{2}\partial_\mu\phi \partial^\mu\phi -
e\epsilon_{\mu\nu}\partial^\nu\phi A^\mu -
\frac{1}{4}F_{\mu\nu}F^{\mu\nu} + \frac{1}{2}m_0 A_{\mu}A^{\mu}.
\label{BLM}
\end{equation}
Here Lorentz indices runs over the two values $0$ and $1$
corresponding to the two space time dimension and the rest of the
notation are standard. The antisymmetric tensor is defined with
the convention $\epsilon_{01}=+1$. The coupling constant $e$ has
one mass dimension in this situation. The parameter $m_0$ is
introduced to represent the masslike term for the gauge field at
the classical level in the same way as it was done in the
Thirring-Wess model \cite{THIR}. In \cite{USHA1, USHA2} the
authors used $m_0=ae^2$. Here we  would like to mention that the
authors in \cite{USHA1} used a regularization parameter $M$ in the
generalized bosonized lagrangian and set it to zero along with few
others parameter to get vector Schwinger mode. Then again the
author added a mass like term for the gauge field ${\cal L}_m =
\frac{1}{2}ae^2A_\mu A^\mu$ and termed this $a$ again as standard
regularization parameter. It is highly confusing at this level.

Let us now proceed to study the phase space structure of the
 model. To this end it is necessary to calculate the
the momenta corresponding to the field $A_0$, $A_1$ and $\phi$.
From the standard definition of the momentum we obtain
\begin{equation}
\pi_0=0 \label{M1},
\end{equation}
\begin{equation}
\pi_1 = F_{01}\label{M2},
\end{equation}
\begin{equation}
\pi_\phi = \dot\phi - eA_1\label{M3},
\end{equation}
where $\pi_0$, $\pi_1$ and $\pi_\phi$ are the momenta
corresponding to the field $A_0$, $A_1$ and $\phi$. Using the
equations (\ref{M1}), (\ref{M2}) and (\ref{M3}), the Hamiltonian
density are calculated:
\begin{equation}
{\cal H} = \frac{1}{ 2}(\pi_\phi +eA_1)^2 + \frac{1}{2}\pi_1^2 +
\frac{1}{2}\phi'^2 + \pi_1A_0' - eA_0\phi' - {1\over 2}m_0(A_0^2 -
A_1^2).\end{equation} Note that $\omega = \pi_0 \approx 0$, is the
familiar primary constraint of the theory. The preservation of the
 constraint $\omega$
requires $[\omega(x), H(y)] = 0$, which leads to the Gauss' law as
a secondary constraint:
\begin{equation}
\tilde\omega = \pi_1' + e\phi' + m_0A_1 \approx 0. \label{SCO}
\end{equation}
The constraints (\ref{M1}) and (\ref{SCO}) form a second class
set. Treating (\ref{M1}) and (\ref{SCO}) as strong condition one
can eliminate $A_0$ and obtain the reduced Hamiltonian density as
follows.
\begin{equation}
{\cal H}_r = \frac{1}{2}(\pi_\phi + eA_1)^2 +
\frac{1}{2m_0}(\pi'_1 + e\phi')^2 + \frac{1}{2}({\pi_1}^2 +
\phi'^2)^2 + \frac{1}{2}m_0A_1^2. \label{RHAM}
\end{equation}
According to the Dirac's prescription \cite{DIR} of quantizing a
theory with second class constraint the Poisson brackets becomes
inadequate for this situation. This type of systems however remain
consistent with the Dirac brackets \cite{DIR}. It is
straightforward to show that the Dirac brackets between the fields
describing the reduced hamiltonian (\ref{RHAM}) remain canonical.
Using the canonical Dirac brackets the following first order
equations of motion  are found out from the reduced Hamiltonian
density (\ref{RHAM}).
\begin{equation}
\dot A_1= \pi_1 -\frac{1}{m_0}(\pi_1'' + e\phi''), \label{EQM1}
\end{equation}
\begin{equation}
\dot\phi = \pi_\phi + eA_1,\label{EQM2}\end{equation}
\begin{equation}
\dot \pi_\phi = (1+ \frac{e^2}{m_0})\phi'' + \frac{e}{m_0}\pi_1''
, \label{EQM3}
\end{equation}
\begin{equation}
\dot\pi_1 = -e\pi_\phi - (m_0+e^2)A_1. \label{EQM4}
\end{equation}
Note that in \cite{USHA1},  all the equations of motion except the
equation  of motion corresponding to equation (\ref{EQM3}) were
identical. In \cite{USHA1}, the calculational mistake started from
that erroneous equation of motion. A little algebra converts the
above first order equations (\ref{EQM1}), (\ref{EQM2}),
(\ref{EQM3}) and (\ref{EQM4}) into the following second order
equations:
\begin{equation}
(\Box + (m_0+ e^2)\pi_1 = 0, \label{SP1}
\end{equation}
\begin{equation}
\Box[\pi_1 + \frac{m_0+e^2}{e}\phi] = 0. \label{SP2}
\end{equation}
Equation (\ref{SP1}) describes a massive boson field with mass $m
=\sqrt{m_0+ e^2}$ whereas equation (\ref{SP2}) describes a
massless scalar field. The result clearly shows that the mass
acquires a generalized expression with the parameter involved in
the masslike term at the classical level as it can be expected
from the result of the paper \cite{AR}. This boson can be
identified with the photon that acquired mass via a dynamical
symmetry breaking. The massless boson (\ref{SP2}) is equivalent to
the massless fermion in (1+1) dimension. So fermion here does not
confine. It remains free as it has been found in chiral Schwinger
model \cite{JR, RABIN, GIR, GIR1, GIR2} and the so called
nonconfining Schwinger model \cite{AR, AR1}. In this context we
should mention that there is some confusion in the literature
regarding the conclusion concerning confinement and de-confinement
scenario of fermion but the result in this context is considered
to be more or less standard now \cite{LO, AR, GIR2, PM, MG, AR1}.
The nature of the theoretical spectrum becomes more transparent if
we calculate the fermionic propagator to which we now tern.

To calculate fermion propagator one needs to work with the
original fermionic model. The calculation is analogous to the so
called nonconfining Schwinger model \cite{AR}. The same
calculation for chiral Schwinger model is available in \cite{GIR,
GIR1}. The effective action obtained by integrating out $\phi$
from the bosonized action (\ref{BLM}) is
\begin{equation}
S_{eff}=\int d^2x\frac{1}{2}[A_\mu(x)
M^{\mu\nu}A_\nu(x)],\end{equation} where,
\begin{equation}
M^{\mu\nu}= m_0 g^{\mu\nu} - \frac{\Box
+e^2}{\Box}\tilde\partial^\mu\tilde\partial^\nu.\end{equation}
Here we have used the standard notation
$\tilde\partial^\mu=\epsilon^{\mu\nu}\partial_\nu.$
 The gauge field propagator is just
the inverse of $M^{\mu\nu}$ and it is found to be
\begin{equation}
\Delta_{\mu\nu}(x-y)=\frac{1}{m_0}[g_{\mu\nu}+ \frac{\Box +e^2}{
\Box(\Box + (m_0+e^2))} \tilde\partial_\mu\tilde\partial_\nu
]\delta(x-y).
\end{equation}
Note that the two poles of propagator are found at the expected
positions. One at zero and another at $m_0 + e^2$ indicating a
massive and a massless excitations.

Setting an  Ansatz, for the Green function of the Dirac operator
$(i\partial\!\!\!\!/ -gA\!\!\!\!/)$, enable us to construct the
propagator of the original fermion $\psi$. The conventional
construction of the Ansatz is
\begin{equation}
G(x,y;A)=e^{ie(\Phi(x)-\Phi(y))}S_F(x-y),\label{ANST}
\end{equation} where $S_F$  is  the  free, massless fermion
propagator and $\Phi$ is determined when the Ansatz (\ref{ANST})
is plugged into the equation for the Green function. From the
standard construction the Green function can be written down as
\begin{equation}
G(x,y;A)=e^{ie\int d^2zA^\mu(z)J_\mu(z)}S_F(x-y),\end{equation}
where the {\it current} $J_\mu$ has the following expression.
\begin{equation}
J_\mu=(\partial^z_\mu  +  \gamma_5\tilde\partial^z_\mu)(D_F(z- x)
-D_F(z-y)).\end{equation} Here $D_F$ is the propagator of a
massless  free  scalar  field. Such propagators have to be
infra-red regularized in two dimensions \cite{LO}:
\begin{equation}
D_F(x)=-{i\over 4\pi}ln(-\mu^2x^2+i0),\end{equation} where $\mu$
is the infra-red regulator mass.

Finally we obtain the  fermion  propagator by functionally
integrating $G(x,y;A)$ over the gauge field:
\begin{eqnarray}
S'_F&=&\int {\cal D}A e^{\frac{i}{2}\int d^2z~(A_\mu(z)
M^{\mu\nu}A_\nu(z) + 2eA_\mu   J^\mu)}S_F(x-y)\nonumber\\&=&{\cal
N} \exp[\frac{D_F}{\frac{m_0^2 + m_0e^2}{e^4}}+
\frac{\Delta_F(m^2=m_0 +
e^2)}{\frac{m_0+e^2}{e^2}}]~~S_F.\label{PROP} \end{eqnarray} Here
$\Delta_F$ is the propagator of a massive free scalar field and
${\cal N}$ is a wave function renormalization factor.

Both the results therefore confirms strongly that the mass
acquires a generalized expression with bare coupling constant and
the parameter involved in the masslike term for the gauge field.
It is known that in vector Schwinger model this type of parameter
free mass generates. In fact, in the bosonized vector Schwinger
model no such parameter exists if gauge fixing is not introduced
at the lagrangian level. In this context, note that setting $a=0$,
in the bosonized lagrangian of the so called nonconfining
Schwinger model \cite{AR} one can obtain vector Schwinger model
\cite{SCH}.

So from our result it can be concluded that it is the
calculational error that led the authors of \cite{USHA1, USHA2} to
land in to this type of untrustworthy result. We have pointed out
in the body of this paper where the the actual error started.
Actually, the use of that erroneous equation corresponding to the
equation  (\ref{EQM3}) led them to reach to the wrong expression
of the mass term for photon. The fermionic propagator presented in
\cite{USHA1, USHA2} also carried the same error. In
equation(\ref{PROP}) of this note the correct expression fermionic
operator is calculated.

As a concluding remark we would like to mention that the term
$\frac{1}{2}ae^2A^\mu A_\mu$ is not a legitimate gauge fixing
term. With this term the vector Schwinger model turns into the so
called nonconfining Schwinger model \cite{AR, AR1}. Introduction
of proper gauge fixing term like $\frac{\alpha}{2e^2}(\partial_\mu
A^\mu)^2$ in the starting lagrangian of vector Schwinger  model
however leads to a result that corresponds to parameter free mass
term for the photon though there exists a parameter $\alpha$ in
the starting lagrangian.


\end{document}